\documentstyle[sprocl]{article}

\def\Journal#1#2#3#4{{#1} {\bf #2}, #3 (#4)}
\def\NPB{{\em Nucl. Phys.} B}
\def\PLB{{\em Phys. Lett.}  B}

\def\PRD{{\em Phys. Rev.} D}

\def\be{\begin{equation}}
\def\ee{\end{equation}}
\def\bea{\begin{eqnarray}}
\def\eea{\end{eqnarray}}

\begin{document}
\title{$\Lambda_b$ LIFETIME FROM THE HQET SUM RULE} 
\author{CHAO-SHANG HUANG$^a$, CHUN LIU$^{a,b}$ and SHI-LIN ZHU$^{c,a}$}
\address{}
\address{$^a$Institute of Theoretical Physics, CAS, Beijing, China\\
         $^b$Institut f\"ur Physik, Universit\"at Mainz, Germany\\ 
         $^c$Kellogg Radiation Laboratory 106-38, 
             California Institute for Technology, USA}

\maketitle\abstracts{ The HQET sum rule analysis for the $\Lambda_b$ 
matrix element of the four-quark operator relevant to its lifetime is 
reported.  Our main conclusion is that the lifetime ratio 
$\tau(\Lambda_b)/\tau(B^0)$ can be as low as $0.91$.}

\section{Introduction}
The experimental result on the lifetime ratio of the $\Lambda_b$ baryon 
and $B$ meson, $\tau(\Lambda_b)/\tau(B^0)=0.795\pm 0.053$ \cite{ex}, still 
needs theoretical understanding.  It can be calculated systematically by 
heavy quark expansion \cite{hqe}, if we do not assume the failure of the 
local duality assumption.  To the order of $1/m_b^2$, the calculated ratio 
is still close to unity.  The potential importance of the $O(1/m_b^3)$ 
effect has been pointed out \cite{inclusive1,inclusive2,history}.  The 
lifetime ratio was calculated as follows \cite{inclusive1},
\bea
\label{1}
\frac{\tau(\Lambda_b)}{\tau(B^0)}&\simeq &0.98
+\xi\{p_1B_1(m_b)+p_2B_2(m_b)+p_3\epsilon_1(m_b)+p_4\epsilon_2(m_b) 
\nonumber\\ &&+[p_5+p_6\tilde{B}(m_b)]r(m_b)\}\; ,
\eea
where the term proportional to $\xi\equiv (f_B/200 {\rm MeV})^2$
arises from the $1/m_b^3$ contributions.  At the scale $m_b$, the values 
of the perturbative coefficients $p_i$'s are $p_1=-0.003$, $p_2=0.004$, 
$p_3=-0.173$, $p_4=-0.195$, $p_5=-0.012$, $p_6=-0.021$.  $B_1$, $B_2$, 
$\epsilon_1$, $\epsilon_2$, $r$ and $\tilde{B}$ are the parameterization 
of the hadronic matrix elements of the following four-quark operators,
\bea
\label{2}
\langle\bar{B}|\bar{b}\gamma_{\mu}(1-\gamma_5)q\bar{q}\gamma^{\mu}
(1-\gamma_5)b|\bar{B}\rangle & \equiv & B_1f_B^2m_B^2 \; , \nonumber\\
\langle\bar{B}|\bar{b}(1-\gamma_5)q\bar{q}(1+\gamma_5)b
|\bar{B}\rangle & \equiv & B_2f_B^2m_B^2 \; , \nonumber\\
\langle\bar{B}|\bar{b}\gamma_{\mu}(1-\gamma_5)t_aq\bar{q}\gamma^{\mu}
(1-\gamma_5)t_ab|\bar{B}\rangle & \equiv & \epsilon_1f_B^2m_B^2 \; ,
\nonumber\\
\langle\bar{B}|\bar{b}(1-\gamma_5)t_aq\bar{q}(1+\gamma_5)t_ab
|\bar{B}\rangle & \equiv & \epsilon_2f_B^2m_B^2 \; ,
\eea
and
\bea
\label{3}
\displaystyle\frac{1}{2m_{\Lambda_b}}\langle\Lambda_b|
\bar{b}\gamma_{\mu}(1-\gamma_5)q\bar{q}\gamma^{\mu}(1-\gamma_5)b
|\Lambda_b\rangle & \equiv & \displaystyle-\frac{f_B^2m_B}{12}r \; ,
\nonumber\\[3mm]
\displaystyle\frac{1}{2m_{\Lambda_b}}\langle\Lambda_b|
\bar{b}(1-\gamma_5)q\bar{q}(1+\gamma_5)b|\Lambda_b\rangle & \equiv &
\displaystyle-\tilde{B}\frac{f_B^2m_B}{24}r \; .
\eea
These parameters have been calculated by QCD sum rules.  In 
Refs. \cite{Blt,Blt1}, the mesonic parameters $B_i$ and $\epsilon_i$ 
were calculated within the framework of heavy quark effective theory 
(HQET).  The baryonic parameters $r$ and $\tilde{B}$ were calculated 
in Refs. \cite{bylt1} and \cite{ours}.  Here we report our result of 
\cite{ours}.

\section{The Calculation}
The new ingredients of our analysis compared to Ref. \cite{bylt1} is the 
following.  (1) Gluon condensate and six-quark condensate are included.  
(2) A different duality assumption is adopted. 

The result of $\tilde{B}=1$ does not change in the valence quark 
approximation.

To calculate $r$, the following three-point Green's function is constructed,
\be
\label{4}
\Pi(\omega, \omega') = i^2\int dxdye^{ik'\cdot x-ik\cdot y}\langle 0|
{\cal T}\tilde{j}^v(x) {\tilde O}(0) \bar{\tilde{j}^v}(y)|0\rangle \; ,
\ee
where $\omega=v\cdot k$ and $\omega'=v\cdot k'$.  The $\Lambda_Q$ baryonic
current $\tilde{j}^v$ is \cite{baryon1,baryon2,baryon3,baryon4},
\be
\label{5}
\tilde{j}^v = \epsilon^{abc}q_1^{Ta}C\gamma_5(a+b\not v)\tau
q_2^b h_v^c \; ,
\ee
where $a$ and $b$ are certain constants, $h_v$ is the heavy quark field in 
the HQET with velocity $v$, $C$ is the charge conjugate matrix, $\tau$ is 
the flavor matrix for $\Lambda_Q$.  In Eq. (\ref{4}), ${\tilde O}$ denotes 
the four-quark operator
\be
\label{6}
{\tilde O} = \bar{h_v}\gamma_{\mu}{1-\gamma_5\over 2}h_v
\bar{q}\gamma^{\mu}{1-\gamma_5\over 2}q \; .
\ee
Note $<\Lambda_b |{\tilde O}|\Lambda_b > =-<\Lambda_b |O|\Lambda_b >$ 
where
\be
\label{7}
O= \bar{h_v}\gamma_{\mu}{1-\gamma_5\over 2}q
\bar{q}\gamma^{\mu}{1-\gamma_5\over 2} h_v \; .
\ee
In terms of the hadronic expression, the parameter $r$ appears in the
ground state contribution of $\Pi(\omega, \omega')$,
\be
\label{8}
\Pi(\omega, \omega') = {1\over 2}\frac{f_{\Lambda}^2\langle\Lambda_Q|O
|\Lambda_Q\rangle}{(\bar{\Lambda}-\omega)(\bar{\Lambda}-\omega')}
\frac{1+\not v}{2} + {\rm higher~~states} \;. 
\ee
$\bar{\Lambda}=m_{\Lambda_Q}-m_Q$ and the quantity $f_{\Lambda}$ is 
defined as $\langle 0|\tilde{j}^v|\Lambda_Q\rangle\equiv f_{\Lambda} u$ 
with $u$ being the unit spinor in the HQET.  The QCD sum rule 
calculations for $f_{\Lambda}$ were given in Refs. 
\cite{baryon1,baryon2,baryon3,baryon4}.  

On the other hand, this Green's function $\Pi(\omega, \omega')$ can be 
calculated in terms of quark and gluon language with vacuum condensate 
straightforwardly.  The fixed point gauge is used \cite{sr1}.  The 
tadpole diagrams in which the light quark lines from the four-quark 
vertex are contracted have been subtracted.  While the calculation can 
be justified if ($-\omega$) and ($-\omega'$) are large, however the 
hadron ground state property should be obtained at small ($-\omega$) 
and ($-\omega'$).  These contradictory requirements are achieved by 
introducing double Borel transformation for $\omega$ and $\omega'$.  

\section{Duality Assumption}
Generally the duality is to simulate the higher states by the whole quark 
and gluon contribution above some threshold energy $\omega_c$.  The whole
contribution of the three-point correlator $\Pi(\omega, \omega')$ can be 
expressed by the dispersion relation,
\be
\label{10}
\Pi(\omega, \omega') = \frac{1}{\pi}\int_0^{\infty}d\nu\int_0^{\infty}
d\nu'\frac{{\rm Im}\Pi(\nu, \nu')}{(\nu-\omega)(\nu'-\omega')} \; .
\ee
With the redefinition of the integral variables
\be
\nu_+  =  \displaystyle\frac{\nu+\nu'}{2} \;, ~~
\nu_-  =  \displaystyle\frac{\nu-\nu'}{2} \; ,
\ee
the integration becomes
\be
\label{12}
\int_0^{\infty}d\nu\int_0^{\infty}d\nu'... =
2\int_0^{\infty}d\nu_+\int_{-\nu_+}^{\nu_+}d\nu_- ... \; .
\ee
It is in $\nu_+$ that the quark-hadron duality is assumed \cite{n,bs},
\be
\label{13}
{\rm higher~~states} = \frac{2}{\pi}\int_{\omega_c}^{\infty}d\nu_+
\int_{-\nu_+}^{\nu_+}d\nu_-
\frac{{\rm Im}\Pi(\nu, \nu')}{(\nu-\omega)(\nu'-\omega')} \; .
\ee
This kind of assumption was suggested in calculating the Isgur-Wise
function in Ref. \cite{n} and was argued for in Ref. \cite{bs}.  

The sum rule for $\langle\Lambda_Q|{\tilde O}|\Lambda_Q\rangle$ after 
the integration with the variable $\nu_-$ is 
\bea
\label{14}
{(a+b)^2\over 2}
f_{\Lambda}^2\exp\left(-\frac{\bar{\Lambda}}{T}\right)
\langle\Lambda_Q|{\tilde O}|\Lambda_Q\rangle & = &
\displaystyle \int_0^{\omega_c}d\nu \exp\left(-\frac{\nu}{T}\right)
\{ \frac{a^2+b^2}{840\pi^6} \nu^8
-\frac{ab}{6\pi^4}\nu^5 \langle\bar{q}q\rangle \nonumber\\[3mm]
&&\displaystyle+\frac{3(a^2+b^2)}{2048\pi^6}\nu^4\langle g^2_sG^2\rangle
+ \frac{5ab}{48\pi^4} m_0^2\langle\bar{q}q\rangle \nu^3 \nonumber\\[3mm]
&&\displaystyle +\kappa_1 \frac{17(a^2+b^2)}{96\pi^2}
\langle\bar{q}q\rangle^2 \nu^2 \}
-\kappa_2 \frac{ab}{144} \langle\bar{q}q\rangle^3 \;,
\eea
where $\kappa_1$, $\kappa_2$ are the parameters used to indicate the 
deviation from the factorization assumption for the four- and six-quark 
condensates. $\kappa_{1,2}=1$ corresponds to the vacuum saturation 
approximation. $\kappa_1 =(3\sim 8)$ is introduced in order to 
include the nonfactorizable contribution and to fit the data \cite{19}. 
There is no discussion of $\kappa_2$ in literature so we use 
$\kappa_2 =1$.  We shall adopt $a=b=1$ \cite{baryon3} in our numerical 
analysis.  The parameters $f_{\Lambda} and {\bar\Lambda}$ were obtained 
by the HQET sum rule analysis of two-point correlator 
\cite{baryon1,baryon2,baryon3,baryon4}.

Our final sum rule is obtained from Eq.(\ref{14}) by dividing that for
$f_\Lambda$.  The value of $\omega_c$ is $(1.2\pm 0.1)$ GeV.  The sum 
rule window is $T=(0.15 - 0.35 )$ GeV.  We obtain for $\kappa_1 =4$
\be
\label{15}
\langle\Lambda_Q|{\tilde O}|\Lambda_Q\rangle = (1.6\pm 0.4)\times
10^{-2} \mbox{GeV}^3 ~~ {\rm or}~~ r = (3.6\pm 0.9) \;.
\ee
By taking $f_B=200$ MeV.  If we use $\kappa_1=1$, we get
\be
\label{17}
\langle\Lambda_Q|{\tilde O}|\Lambda_Q\rangle = (5.5\pm 1.0)\times
10^{-3} \mbox{GeV}^3 ~~ {\rm or}~~ r = (1.3\pm 0.3) \;.
\ee
Note that our results depend on $\omega_c$ weakly.

The value of $r$ we have obtained above is at some hadronic scale, 
because we have been working in the HQET.  By choosing 
$\alpha_s(\mu_{had})=0.5$ (corresponding to $\mu_{had}\sim 0.67$ GeV), 
we obtain $\tilde{B}(m_b)\simeq 0.58$ and 
\be
r(m_b)\simeq (6.2\pm 1.6)~~{\rm for} ~~\kappa_1=4\;,~~{\rm and}~~
r(m_b)\simeq (2.3\pm 0.6)~~{\rm for} ~~\kappa_1=1\;.
\ee

The $\Lambda_b$ and $B^0$ lifetime ratio given in Eq. (\ref{1}) is 
expressed specifically as
\bea
\frac{\tau(\Lambda_b)}{\tau(B^0)} &\simeq 0.83\pm 0.04 
~~{\rm for} ~~\kappa_1=4\;,\nonumber\\ 
&\simeq 0.93\pm 0.02 ~~{\rm for}~~\kappa_1=1\;.
\eea
Where the values $\epsilon_1(m_b)=-0.08$ and $\epsilon_2(m_b)=-0.01$ 
have been taken from the QCD sum rules \cite{Blt}.  We see that with the 
vacuum saturation ($\kappa_1=1$), although $r$ is enhanced by about six 
times compared to that in Ref. \cite{bylt1}, it is still not large enough 
to account for the data.  The life time ratio between $\Lambda_b $ and 
$B$ mesons can be explained if we also take into account the 
nonfactorizable contribution of the four-quark condensate.

\section{Conclusion}
In summary, we have reanalyzed the QCD sum rule for the $\Lambda_b$ matrix
element of the four-quark operator relevant to the lifetime of $\Lambda_b$.
The difference between Ref. \cite{bylt1} and ours is mainly because of 
duality assumptions.  While a large nonfactorizable effect in the 
four-quark condensate can make the theoretical result consistent with the 
experiment, our main conclusion is that the lifetime ratio 
$\tau(\Lambda_b)/\tau(B^0)$ can be as low as $0.91$.

\section*{Acknowledgments}
This work was supported in part by the National Natural Science Foundation 
of China. C.L. is also supported by the Alexander von Humboldt Foundation.

\end{document}